\documentclass[amsmath,amssymb,aps,10pt,prd,letterpaper,nofootinbib,balancelastpage,notitlepage,superscriptaddress,twocolumn,floatfix,preprintnumbers]{revtex4-2}
\usepackage{graphicx}	
\usepackage{ragged2e}	
\usepackage{bm}		    
\usepackage[sort&compress]{natbib}	 	
	\setcitestyle{square,numbers,comma}	
\usepackage[colorlinks=true,urlcolor=blue,linkcolor=black,citecolor=red]{hyperref}	
\newcommand{\figref}[2][]{Fig{#1}.~\ref{fig:#2}}		
\newcommand{\secref}[2][]{Sec{#1}.~\ref{sec:#2}}		
\newcommand{\appref}[2][x]{Appendi{#1}~\ref{app:#2}}	
\renewcommand{\eqref}[2][]{Eq{#1}.~(\ref{eq:#2})}		
\newcommand{\citeR}[2][]{Ref{#1}.~\cite{#2}}			
\newcommand{\orcid}[1]{\href{https://orcid.org/#1}{\,\includegraphics[width=8px]{ORCID.png}}}
\newcommand{\nl}{\nonumber \\ & \quad }					
\newcommand{\DM}{\mathrm{DM}}
\newcommand{\Hz}{\,\textrm{Hz}}
\newcommand{\Jeff}{\bm J_\mathrm{eff}}
\newcommand{\LL}{\mathcal{L}}
\DeclareMathOperator{\IM}{\mathrm{Im}}
\DeclareMathOperator{\RE}{\mathrm{Re}}
\newcommand{\nhat}{\bm{\hat{n}}}

\newcommand{\Y}{\bm{\hat{y}}}

\begin{document}

\title{Curl up with a good \texorpdfstring{$\bm B$}{B}:\\
Detecting ultralight dark matter with differential magnetometry}
\date{\today}
\author{Itay M. Bloch}
\email{itayblochm@berkeley.edu}
\affiliation{Berkeley Center for Theoretical Physics, University of California, Berkeley, CA 94720, USA}
\affiliation{Theory Group, Lawrence Berkeley National Laboratory, Berkeley, CA 94720, USA}
\author{Saarik Kalia}
\email{kalias@umn.edu}
\affiliation{School of Physics \& Astronomy, University of Minnesota, Minneapolis, MN 55455, USA}

\begin{abstract}
Ultralight dark matter (such as kinetically mixed dark-photon dark matter or axionlike dark matter) can source an oscillating magnetic-field signal at the Earth's surface, which can be measured by a synchronized array of ground-based magnetometers.  The global signal of ultralight dark matter can be robustly predicted for low masses, when the wavelength of the dark matter is larger than the radius of the Earth, $\lambda_\DM\gg R$.  However, at higher masses, environmental effects, such as the Schumann resonances, can become relevant, making the global magnetic-field signal $\bm B$ difficult to reliably model.  In this work, we show that $\nabla\times\bm B$ is robust to global environmental details, and instead only depends on the local dark matter amplitude.  We therefore propose to measure the local curl of the magnetic field at the Earth's surface, as a means for detecting ultralight dark matter with $\lambda_\DM\lesssim R$.  As this measurement requires vertical gradients, it can be done near a hill/mountain.  Our measurement scheme not only allows for a robust prediction, but also acts as a background rejection scheme for external noise sources.  We show that our technique can be the most sensitive terrestrial probe of dark-photon dark matter for frequencies $10\,\mathrm{Hz}\leq f_{A'}\leq1\,\mathrm{kHz}$ (corresponding to masses $4\times10^{-14}\,\mathrm{eV}\leq m_{A'}\leq4\times10^{-12}\,\mathrm{eV}$).  It can also achieve sensitivities to axionlike dark matter comparabe to the CAST helioscope, in the same frequency range.
\end{abstract}

\maketitle
\tableofcontents

\section{Introduction}
\label{sec:introduction}

One of the most prominent open problems in fundamental physics is understanding the nature of dark matter (DM).  A plethora of DM candidates have been proposed, spanning a wide range of masses, but one class of candidates which has gained signficant interest in recent years is ultralight bosonic DM~\cite{Arias:2012az,kimball2022search}.  This class includes bosonic DM candidates with masses $\lesssim1\,\mathrm{eV}$, which have correspondingly large occupation numbers, and so behave like classical fields~\cite{lin2018self,Centers:2019dyn}.  Some of the most popular ultralight DM candidates are QCD axions~\cite{Preskill:1982cy,Abbott:1982af,Dine:1982ah}, axionlike particles~\cite{Svrcek:2006yi,Arvanitaki:2009fg,Gra15,co2020predictions}, and dark photons~\cite{Holdom:1986ag,cvetivc1996implications,Nelson:2011sf,Graham:2015rva}.

All of these candidates may possess couplings to electromagnetism~\cite{Holdom:1986ag,Sikivie:1983ip}, which can be searched for in a variety of laboratory experiments~\cite{ehret2010new,Wagner:2010mi,Redondo:2010dp,Horns:2012jf,Betz_2013,Graham:2014sha,Chaudhuri:2014dla,graham2016dark,Caldwell:2016dcw,Anastassopoulos:2017ftl,Baryakhtar:2018doz,armengaud2019physics,Lawson:2019brd,gramolin2021search,Andrianavalomahefa:2020ucg,Cantatore:2020obc,Salemi:2021gck,Su:2021jvk,Chiles_2022,haystaccollaboration2023new,romanenko2023new}.  The signals in these laboratory experiments typically scale with some power of the characteristic size of the experiment $L$.  In particular, in the DPDM case, this scaling typically appears as $m_{A'}L$ (so long as $m_{A'}L\ll1$), where $m_{A'}$ is the mass of the DPDM~\cite{Chaudhuri:2014dla}.  For laboratory experiments of size $L\sim\mathrm{few~meters}$, searching for DPDM therefore becomes increasingly challenging at masses $m_{A'}\lesssim10^{-7}\,\mathrm{eV}$.

It was recently shown that the Earth can act as a transducer to convert both dark-photon dark matter (DPDM)~\cite{Fedderke:2021rys} and axion%
\footnote{Henceforth, we simply use ``axion" to refer to both the QCD axion and axionlike particles.}
DM~\cite{Arza_2022} into an oscillating magnetic-field signal at the Earth's surface.  The characteristic size $L$ for this effect is the radius of the Earth $R\approx6400\,\mathrm{km}$, which yields a larger signal and allows access to the low-mass DPDM parameter space.  It is also argued in \citeR[s]{Fedderke:2021rys,Arza_2022} that, so long as $m_\DM R\ll1$,%
\footnote{In contexts applying only to dark photons or only to axions, we use $m_{A'}$ or $m_a$, respectively, to denote the DM mass.  In contexts applying to both, we simply use $m_\DM$.}
this signal can be made robust to details of the near-Earth environment, so that it does not depend on complicated atmospheric/geological modeling.

The Earth-transducer signal can be searched for using an array of unshielded ground-based magnetometers.  In \citeR[s]{Arza_2022,Fedderke:2021iqw}, publicly available data from an existing global array of magnetometers, maintained by the SuperMAG Collaboration~\cite{SuperMAGwebsite,Gjerloev:2009wsd}, were used to set limits on axion DM and DPDM parameter space for $2\times10^{-18}\,\mathrm{eV}\lesssim m_\DM\lesssim7\times10^{-17}\,\mathrm{eV}$.  More recently, in July 2022, the SNIPE Hunt Collaboration took simultaneous magnetometer measurements at three radio-quiet locations to constrain parameter space in the higher mass range $2\times10^{-15}\,\mathrm{eV}\lesssim m_\DM\lesssim2\times10^{-14}\,\mathrm{eV}$~\cite{sulai2023hunt}.  Their constraints were limited by magnetometer sensitivity rather than environmental noise.  The collaboration intends to continue its efforts using more sensitive magnetometers in 2023 and 2024.

In this work, we focus on extending the measurement of this effect to even higher DM masses, for which $m_\DM R\gtrsim1$.  In this case, the robustness arguments of \citeR[s]{Fedderke:2021rys,Arza_2022} are no longer valid, and enviromental effects, such as the Schumann resonances~\cite{Schumann+1952+149+154,bliokh1980schumann,Rodriguez-Camacho} can affect the magnetic field signal $\bm B$.  While $\bm B$ itself is affected by these environmental details, $\nabla\times\bm B$ is not, as it is related directly to the local DM amplitude via the Amp\`ere-Maxwell law.  This work therefore proposes to measure the local curl of the magnetic field, in order to obtain a robust measurement of the DM at higher masses.  Such a measurement requires vertical gradients (as will be shown later), and so can be obtained near a hill/mountain.  Measuring gradients of $\bm B$ is more difficult than measuring $\bm B$ itself, and so our scheme naively suffers from a reduced signal.  However because we generically expect environmental magnetic fields to satisfy $\nabla\times\bm B=0$, our measurement scheme should also have reduced backgrounds!  Therefore, this method can still achieve good sensitivity to ultralight DM. The SNIPE Hunt Collaboration intends to implement this scheme in various radio-quiet locations in their 2024 experimental run.

This work is structured as follows.  In \secref{motivation}, we review the magnetic-field signal of ultralight DM derived in \citeR[s]{Fedderke:2021rys,Arza_2022}.  In particular, we pay special attention to the robustness argument for the low-frequency signal, and show why it breaks down at higher frequencies.  We then show how $\nabla\times\bm B$ can alternatively be measured to robustly detect ultralight DM.  In \secref{measurement}, we demonstrate how to practically measure $\nabla\times\bm B$ (see \figref{curlfig} for illustration).  We then compute the sensitivity of this scheme to axion DM and DPDM, as shown in \figref{sensitivity}, assuming that the sensitivity remains limited by internal magnetometer noise.  We also outline several other potential sources of noise, and benchmarks that must be met in order for them to remain subdominant.  In \secref{conclusion}, we conclude.  In \appref{contributions}, we compute various contributions to our curl measurement which appear in \secref{measurement}.

\section{Motivation}
\label{sec:motivation}

In this section, we review the magnetic-field signal induced by ultralight DM at the Earth's surface, and explain why at high frequencies, we must measure $\nabla\times\bm B$ to robustly detect it.  This signal was first described for DPDM in \cite{Fedderke:2021rys}, while the corresponding signal for axion DM was first described in \cite{Arza_2022}.  We begin by briefly reviewing these results.  In particular, we explain how the boundary-dependent contributions to the signals can be projected out at low frequencies, leaving only the contributions which are insensitive to environmental details.  We then move on to explain why this projection scheme fails at frequencies where the Compton wavelength of the DM is comparable to or smaller than the radius of the Earth $R$.  We outline how the dependence on environmental details can instead be eliminated by measuring the component of $\nabla\times\bm B$ which is parallel to the Earth's surface.

\subsection{Review of signal}
\label{sec:lowf}

In this work, we consider two models of ultralight DM, both of which can induce a magnetic-field signal at the Earth's surface through their respective couplings to electromagnetism.  The first is a kinetically mixed dark photon $A'$, with mass $m_{A'}=2\pi f_{A'}$ and kinetic mixing parameter $\varepsilon\ll1$.  The Lagrangian describing the interaction between the dark photon and Standard Model (SM) photon is given by%
\footnote{The Lagrangian for the mixed photon--dark-photon system can be written in multiple different bases (see Sec.~II\,A and Appendix~A of \citeR{Fedderke:2021rys} for a detailed review).  In this work, we operate only in the so-called ``interaction basis," in which the Lagrangian is given by \eqref{DPDM_Lagrangian}.  In this basis, only $A$ interacts with SM currents at leading order.  However, $A$ and $A'$ are not propagation eigenstates, and so will mix as they propagate through vacuum.}
\begin{align}
    \LL_{A'}&\supset-\frac14F_{\mu\nu}F^{\mu\nu}-\frac14F'_{\mu\nu}F'^{\mu\nu}+\frac12m_{A'}A'_\mu A'^\mu\nl
    +\varepsilon m_{A'}^2A_\mu A'^\mu-J_\mathrm{EM}^\mu A_\mu.
    \label{eq:DPDM_Lagrangian}
\end{align}
Here $F'_{\mu\nu}=\partial_\mu A'_\nu-\partial_\nu A'_\mu$ is the field-strength tensor for the dark photon, and $J^\mu_\mathrm{EM}$ is the SM electromagnetic current.

By comparing the last two terms of \eqref{DPDM_Lagrangian}, it can be seen that in the $\varepsilon\ll1$ limit where backreaction can be neglected, $A'$ plays a role equivalent to that of an electromagnetic current.  In particular, in this limit, we can treat $A'$ as nondynamical and parametrize its effect entirely by an ``effective background current"
\begin{equation}
    J^\mu_\mathrm{eff}=-\varepsilon m_{A'}^2A'^\mu.
    \label{eq:DPDM_current}
\end{equation}
(See Appendix~A\,3 of \citeR{Fedderke:2021rys} for a detailed justification of the effective current approach.)  As we will be considering $A'$ to constitute the DM, we will be primarily interested in the case where it is non-relativistic, $v_\DM\ll1$.  The equations of motion implied by \eqref{DPDM_Lagrangian} enforce $\partial_\mu A'^\mu=0$, so in the non-relativistic case, $A'^0=0$ and \eqref{DPDM_current} has only spatial components (i.e., there is no associated ``effective charge").

The second model we consider is that of an axion $a$, with mass $m_a=2\pi f_a$ and axion-photon coupling $g_{a\gamma}$.%
\footnote{Note that the frequency $f_a$ should not be confused with the axion decay constant, which is inversely proportional to $g_{a\gamma}$.}
Its interaction with electromagnetism is described by the Lagrangian
\begin{equation}
    \LL_a\supset\frac12\partial_\mu a\partial^\mu a-\frac14F_{\mu\nu}F^{\mu\nu}-\frac12m_aa^2+\frac14g_{a\gamma}aF_{\mu\nu}\tilde F^{\mu\nu},
    \label{eq:ax_Lagrangian}
\end{equation}
where $\tilde F^{\mu\nu}=\frac12\epsilon^{\mu\nu\rho\sigma}F_{\rho\sigma}$.  Conveniently, in the non-relativistic limit, the effect of axion DM on electromagnetism can also be parametrized entirely by a background current (see Sec.~II\,A of \citeR{Arza_2022})
\begin{equation}
    \Jeff=-g_{a\gamma}(\partial_ta)\bm B.
    \label{eq:ax_current}
\end{equation}
While the dark photon current in \eqref{DPDM_current} depends only on the presence of DPDM, the axion current in \eqref{ax_current} depends on both the presence of axion DM and a nonzero static magnetic field.  In our case, the Earth's DC magnetic field $\bm B_\oplus$ will play this role in \eqref{ax_current}.

The effects of both axion DM and DPDM can thus be described in a common framework, which relates the DM amplitude to an effective current.  In the case where the de Broglie wavelength of the DM is much larger than the radius of the Earth,
\begin{equation}
     \lambda_\mathrm{dB}=v_\DM^{-1}\lambda_\DM\sim(v_\DM f_\DM)^{-1}\gg R,
\end{equation}
spatial variations in the DM amplitude can be neglected on lengthscales comparable to $R$.  Therefore the DM can be treated as a spatially uniform classical field oscillating at angular frequency $m_\DM$.  In the axion case,
\begin{equation}
    a(\bm x,t)=\RE\left[a_0e^{-im_at}\right]
    \label{eq:ax_wave}
\end{equation}
for some complex amplitude $a_0$, while in the dark photon case
\begin{equation}
    \bm A'(\bm x,t)=\RE\left[\sum_{m=-1}^1A'_{0,m}\nhat_me^{-im_{A'}t}\right]
    \label{eq:DPDM_wave}
\end{equation}
for complex amplitudes $A'_{0,m}$ (which may have different phases for each $m$).  Here the unit vectors $\nhat_m$ denote the three helicity polarizations of the dark photon
\begin{align}
    \nhat_\pm&=\mp\frac1{\sqrt2}(\bm{\hat x}\pm i\bm{\hat y}),\\
    \nhat_0&=\bm{\hat z}.
\end{align}

Note that \eqref[s]{ax_wave} and (\ref{eq:DPDM_wave}) are only valid within a coherence time $T_\mathrm{coh}\sim2\pi/(m_\DM v_\DM^2)$.  Over timescales longer than this, the DM amplitude will vary, and will be essentially random from one coherence time to the next.  The distribution of DM amplitudes is normalized by
\begin{equation}
    \frac12m_a\langle|a_0|^2\rangle=\frac12m_{A'}\sum_m\langle|A'_{0,m}|^2\rangle=\rho_\DM,
\end{equation}
where $\rho_\DM\approx0.3\,\mathrm{GeV/cm}^3$ is the local DM density, and $\langle\cdots\rangle$ denotes the expectation over many coherence times.

As the effective current $\Jeff$ is related to the DM amplitude, it will inherit the coherence properties of the DM.  Namely, in the DPDM case, $\Jeff$ should be spatially uniform (on $\mathcal O(R)$ lengthscales) and monochromatic with frequency $m_{A'}$, on timescales shorter than $T_\mathrm{coh}$.%
\footnote{\label{ftnt:frames}%
Note that in the DPDM case, $\Jeff$ maintains a fixed orientation in the \emph{non-rotating celestial frame}.  If the measurement apparatus is fixed to the Earth, this means that $\Jeff$ precesses in the rotating frame of the experiment.  This leads to the $f_d$-dependence in \eqref{DPDM_signal}.  In contrast, the direction of $\Jeff$ in the axion case is set by $\bm B_\oplus$, which co-rotates with the Earth.  There is thus no $f_d$-dependence in \eqref{ax_signal}.}
On longer timescales, the effective current can change both amplitude and direction.  In the axion DM case, however, the direction of $\Jeff$ is set by $\bm B_\oplus$.  Since the Earth's magnetic field is (approximately) dipolar, then $\Jeff$ will not be uniform in space.  Additionally, since $a$ only sets the amplitude, not the direction of $\Jeff$, then on timescales longer than $T_\mathrm{coh}$, only the amplitude of $\Jeff$ will vary along with the DM amplitude.  The direction of $\Jeff$ instead varies on the (much longer) timescale over which the Earth's magnetic field drifts.

Once the DM has been parametrized in this effective current framework, the magnetic field signal at the Earth's surface simply comes from solving Maxwell's equations with a nonzero background current.  In order to do so, the boundary conditions of the system must be specified.  The simplest scenario invoked in \citeR[s]{Fedderke:2021rys,Arza_2022} took the relevant boundaries of the system to be the surface of the Earth (of radius $R$) and the ionosphere (of radius $R+h$, with $h\sim100\,\mathrm{km}\ll R$), which were treated as perfectly conducting concentric spheres, as shown in \figref{earthfig}.  Once these boundary conditions have been specified, it is straightforward to solve for the effect of the DPDM current \eqref{DPDM_current} [with the form of $\bm A'$ in \eqref{DPDM_wave}].  In the $m_{A'}h\ll1$ limit, the resulting magnetic field at the Earth's surface is given by~\cite{sulai2023hunt}
\begin{align}
    \bm B_{A'}(\Omega,t) &= \sqrt{\frac{4\pi}{3}}\cdot\frac{m_{A'}R}{2-(m_{A'}R)^2}\cdot    \varepsilon m_{A'}\nl
    \cdot\mathrm{Re}\left[ \sum_{m=-1}^{1} A_m^{'} \mathbf{\Phi}_{1m}(\Omega) e^{-2 \pi i(f_{A'}-f_dm)t}\right].
    \label{eq:DPDM_signal}
\end{align}
Here $\Omega$ denotes the geographic location on Earth's surface, $f_d=(\text{sidereal day})^{-1}$ [which appears due to the rotation of the Earth; see footnote~\ref{ftnt:frames}], and $\bm\Phi_{\ell m}$ denotes one of the vector spherical harmonics (VSH; see Appendix D of \citeR{Fedderke:2021rys}).

In the axion DM case, the only additional ingredient that is required is a model for the Earth's magnetic field.  \citeR{Arza_2022} utilized the IGRF-13 model~\cite{IGRF}, which parametrizes the Earth's magnetic field in terms of a scalar potential $\bm B_\oplus=-\nabla V_0$ that is expanded as
\begin{equation}
    V_0=\sum_{\ell=1}^\infty\sum_{m=0}^\ell\frac{R^{\ell+2}}{r^{\ell+1}}(g_{\ell m}\cos(m\phi)+h_{\ell m}\sin(m\phi))P_\ell^m(\cos\theta),
\end{equation}
where $P_\ell^m$ are the Schmidt-normalized associated Legendre polynomials, and $g_{\ell m}$ and $h_{\ell m}$ are the Gauss coefficients specified by the IGRF model (see Table 2 of \citeR{IGRF}).  The IGRF model for $\bm B_\oplus$ and \eqref{ax_wave} then fully determine the axion effective current in \eqref{ax_current}.  If the above-mentioned model of the Earth-ionosphere cavity is again assumed, then the magnetic field at the Earth's surface, sourced by the axion effective current, is given by~\cite{sulai2023hunt}
\begin{equation}
    \bm B_a=g_{a\gamma}a_0\cdot\IM\left[\sum_{\ell m}\frac{(\ell+1)C_{\ell m}m_aR}{\ell(\ell+1)-(m_aR)^2}\cdot\bm\Phi_{\ell m}e^{-2\pi if_at}\right],
    \label{eq:ax_signal}
\end{equation}
where
\begin{equation}
    C_{\ell m}=(-1)^m\sqrt{\frac{4\pi(2-\delta_{m0})}{2\ell+1}}\frac{g_{\ell m}-ih_{\ell m}}2.
\end{equation}

\subsection{Robustness of low-frequency signal}
\label{sec:robustness}

The signals described by \eqref[s]{DPDM_signal} and (\ref{eq:ax_signal}) can depend on the boundary conditions of the near-Earth environment.  It is argued in Sec.~II\,B of \citeR{Fedderke:2021rys} that the perfectly conducting spherical model of the Earth-ionosphere cavity is valid for frequencies $(\mathrm{few})\times10^{-16}\,\mathrm{eV}\lesssim\omega\lesssim R^{-1}$.  Since \citeR[s]{Fedderke:2021rys,Arza_2022} were interested in DM masses below this range, though, they employed a crucial mathematical result in their analysis, namely that the boundary-dependent contributions to the magnetic field signal appear at leading order in different VSH components as compared to the contributions in \eqref[s]{DPDM_signal} and (\ref{eq:ax_signal}).  The practical implication was that by projecting the measured global magnetic field signal onto the appropriate VSH components, any boundary-dependent effects can effectively be projected out, leaving only the contributions in \eqref[s]{DPDM_signal} and (\ref{eq:ax_signal}).  By performing this projection, they could isolate a signal which was robust to environmental details, even at low frequencies where the Earth and ionosphere fail to behave exactly like perfectly conducting spheres.

The argument justifying the projection relies on showing that the electric field in the lower atmosphere vanishes to leading order (see, for example, Appendix B.2. of \citeR{Arza_2022}).  The main idea of the argument is that since the electric field must vanish sufficiently deep within the Earth's crust and sufficiently far above the Earth's surface, if the DM Compton wavelength is much larger than the characteristic size of the nonconducting lower atmosphere, it should vanish to leading order everywhere, regardless of the details of the conductivity profile in/around the Earth. To understand the key points of the argument in more detail, let us no longer assume the spherical Earth-ionosphere cavity model, but let us still assume that the near-Earth environment can be described by a cavity of size $\mathcal O(R)$.  That is, let us assume that there are some inner and outer boundaries $\Sigma_1$ and $\Sigma_2$, each of characteristic radius $\sim R$, with a distance of characteristic length $\sim h\ll R$ between them, so that $\bm E_\parallel|_{\Sigma_1,\Sigma_2}=0$.%
\footnote{The region between $\Sigma_1$ and $\Sigma_2$ may not be entirely vacuum.  For instance, we may imagine $\Sigma_1$ to lie in the upper mantle and $\Sigma_2$ to lie deep in the ionosphere, so that the region between them contains the crust, lower atmosphere, and parts of the ionosphere.  All these regions have nonzero conductivity, but the crust and lower ionosphere have conductivities $\sigma\gtrsim10^{-4}\,\mathrm{S/m}$~\cite{atlas,Takeda:1985hcf}, which correspond to skin depths $\delta\sim\sqrt{2/\sigma m_\DM}\lesssim50\,\mathrm{km}$ (for $f_\DM\gtrsim1\,\mathrm{Hz}$).  This means we can take $\Sigma_1$ and $\Sigma_2$ to be $\mathcal O(50)\,\mathrm{km}$ below the Earth's surface and deep into the ionosphere, respectively, while ensuring that $\bm E_\parallel|_{\Sigma_1,\Sigma_2}=0$.  Then the distance between them is $h\sim100\,\mathrm{km}\ll R$.}
Then in the case that $m_\DM R\ll1$, the entire region between $\Sigma_1$ and $\Sigma_2$ is smaller than a wavelength and so $\bm E$ should be suppressed everywhere inside this cavity.

\begin{figure}[t]
\includegraphics[width=0.99\columnwidth]{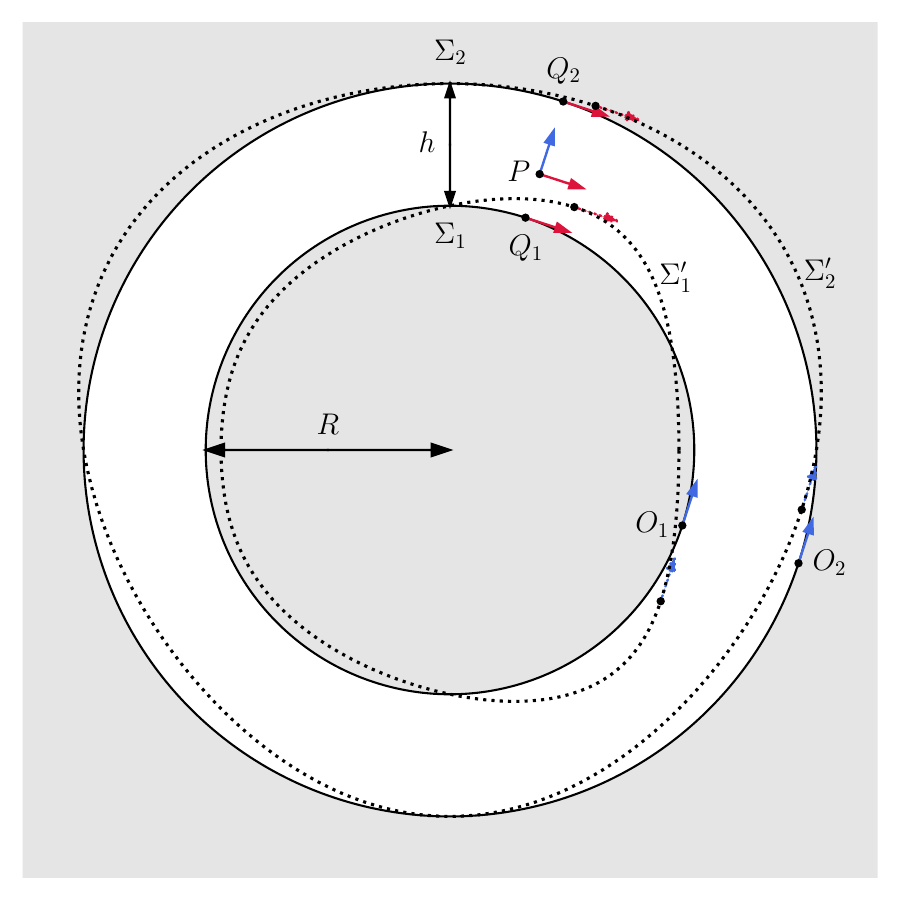}
\caption{\label{fig:earthfig}%
    Diagram of Earth-ionosphere cavity and argument of \secref{robustness}.  The surfaces $\Sigma_1,\Sigma_2$ designate the boundaries of the Earth and ionosphere (scale exaggerated for display), in the idealized spherical model.  The Earth has radius $R$, while the ionosphere has radius $R+h$ (with $h\ll R$).  The surfaces $\Sigma'_1,\Sigma'_2$ show possible boundaries in an non-idealized scenario, in which case they are generically defined as the surfaces where $\bm E_\parallel=0$.  The point $P$ represents a generic point in the interior of the cavity, with tangential (red) and radial (blue) directions shown.  The points $Q_1,Q_2$ are nearby points on $\Sigma_1,\Sigma_2$ whose parallel directions on the surfaces match the tangential directions of $P$.  If $\lambda_\DM\gg h$, then $P,Q_1,Q_2$ are all less than a wavelength apart, and so the tangential electric field at $P$ must be $\mathcal O((h/\lambda_\DM)^2)$.  The points $O_1,O_2$ are far away points on $\Sigma_1,\Sigma_2$, which have one parallel direction matching the radial direction of $P$.  If $\lambda_\DM\gg R$, then additionally, the radial electric field at $P$ must be $\mathcal O((R/\lambda_\DM)^2)$.
    }
\end{figure}

More specifically, consider a generic point $P$ in the region between $\Sigma_1$ and $\Sigma_2$ (see \figref{earthfig}).  Because each component of $\bm E$ satisfies the wave equation separately, these components propagate independently through the cavity.  Therefore, we can evaluate the effect of the boundary conditions on the electric field at $P$, by treating orthogonal directions separately.  First, let us consider the tangential directions. There are nearby points $Q_1,Q_2$ on $\Sigma_1,\Sigma_2$ (where $\bm E_\parallel=0$), respectively, whose parallel directions to the surfaces are the same as the tangential directions of $P$.%
\footnote{Note that in the generic case where $\Sigma_1,\Sigma_2$ are not exactly spherical (e.g., see $\Sigma'_1,\Sigma'_2$ in \figref{earthfig}), there is some ambiguity as to what are the ``tangential" and ``radial" directions at $P$.  The direction which we define as tangential can be varied, so long as $Q_1,Q_2$ remain a distance $\mathcal O(h)$ away from $P$.  Generically, this allows us to vary the tangential direction by an angle of up to $\sim h/R$.  Doing so will mix the tangential and radial directions slightly.  The above argument makes the case that the tangential field is $\mathcal O((m_\DM h)^2)$, while the radial field is $\mathcal O((m_\DM R)^2)$.  Mixing caused from redefining the directions at $P$ can give an additional contribution to the tangential field of order $(m_\DM R)^2\cdot h/R\sim m_\DM^2hR$.  This is still higher order compared to the magnetic field, which is $\mathcal O(m_\DM R)$, and so the electric field can still be neglected.}
These points are each within $\mathcal O(h)$ of $P$, and since the electric field should only vary on wavelength scales $\sim m_\DM^{-1}$, then the tangential electric field at $P$ must be $\mathcal O((m_\DM h)^2)$.  To evaluate the radial direction, instead consider far away points $O_1,O_2$ on $\Sigma_1,\Sigma_2$, respectively, for which the radial direction of $P$ is one of the parallel directions of each of $O_1,O_2$.  Since $O_1,O_2$ are a distance $\mathcal O(R)$ from $P$, then the radial electric field at $P$ should be $\mathcal O((m_\DM R)^2)$.  From \eqref[s]{DPDM_signal} and (\ref{eq:ax_signal}), we can see that the magnetic field is $\mathcal O(m_\DM R)$, and so the electric field in both directions is higher order.

The argument then follows from the Amp\`ere-Maxwell law with an effective current
\begin{equation}
    \nabla\times\bm B-\partial_t\bm E=\Jeff.
    \label{eq:ampere}
\end{equation}
As argued above, if $m_\DM R\ll1$, the electric field is higher order and so the second term in \eqref{ampere} can be neglected.  Therefore, since boundary conditions cannot change $\Jeff$, any change in the boundary conditions of the system can only modify the solution for $\bm B$ by some curl-free contribution.  That is, regardless of boundary conditions,
\begin{equation}
    \bm B=\bm B_\mathrm{sph}+\nabla V,
\end{equation}
where $\bm B_\mathrm{sph}$ are the results \eqref[s]{DPDM_signal} and (\ref{eq:ax_signal}) calculated in the spherical model (which satsify $\nabla\times\bm B_\mathrm{sph}=\Jeff$) and $V$ is some scalar function (so that $\nabla V$ is curl-free).  These two contributions are composed of different types of VSH.  Namely, $\nabla V$ is composed entirely of $\bm Y_{\ell m}$ and $\bm\Psi_{\ell m}$ modes (see Eq.~(A9) of \citeR{Arza_2022}), while $\bm B_\mathrm{sph}$ is composed of $\bm\Phi_{\ell m}$ modes.  Thus by projecting onto only the $\bm\Phi_{\ell m}$ modes, one can isolate the boundary-independent contributions to the signal!

\subsection{Strategy at higher frequencies}
\label{sec:highf}

In this work, we will primarily be interested in detecting DM in the frequency range $10\Hz\leq f_\DM\leq1\,\mathrm{kHz}$, for which $h\lesssim m_\DM^{-1}<R$.  In this frequency range, the argument of \secref{robustness} breaks down.  Even in the idealized spherical case, the radial electric field becomes comparable to the magnetic field solutions in \eqref[s]{DPDM_signal} and (\ref{eq:ax_signal}).  Therefore the second term in \eqref{ampere} cannot be neglected, and even the $\bm\Phi_{\ell m}$ modes of the magnetic-field signal may exhibit boundary dependence.

The impact of environmental effects on the signal can be seen by closely examining the resonance structure of the signals \eqref[s]{DPDM_signal} and (\ref{eq:ax_signal}).  Note that \eqref{ax_signal} diverges at the resonances $m_aR=\sqrt{\ell(\ell+1)}$.%
\footnote{Only the $\ell=1$ resonance is apparent in \eqref{DPDM_signal}.  This is because this signal is calculated in the exact non-relavistic limit $v=0$.  If the DPDM is taken to have nonzero velocity, then the effective current \eqref{DPDM_current} has a mild spatial dependence, which introduces higher $\ell$ modes.  The higher Schumann resonances will then be seen in these higher $\ell$ contributions.}
These are well-studied resonances of the Earth-ionosphere cavity, known as the Schumann resonances~\cite{Schumann+1952+149+154,bliokh1980schumann,Rodriguez-Camacho}.  The idealized spherical model used to calculate \eqref[s]{DPDM_signal} and (\ref{eq:ax_signal}) predicts the first three Schumann resonances at $f=10.6\Hz,18.3\Hz,25.9\Hz$.  Empirically, the Schumann resonances exhibit significant diurnal and seasonal variations, but the central frequencies of the first three resonances take typical values of 7.4--8.0\Hz, 13.7--14.6\Hz, and 19.9--21.1\Hz~\cite{Rodriguez-Camacho}.  The measured widths of these peaks also vary significantly, but often are as low as 1.5--2\Hz~\cite{Rodriguez-Camacho}.  This means the spherical model significantly mis-predicts the actual resonance structure of the signal, potentially leading to order-of-magnitude errors in the predicted signal.  For instance at precisely $f_\DM=10.6\Hz$, the signals \eqref[s]{DPDM_signal} and (\ref{eq:ax_signal}) formally predict an infinite magnetic-field signal, when in fact, at many times of day/year, there is likely no resonant enhancement at all because this frequency lies outside the first Schumann resonance.

We therefore require a different strategy at higher frequencies, which will be robust to enviromental effects on the signal, particularly the Schumann resonances.  We note that the Schumann resonances are not present in the effective current $\Jeff$ itself.  They arise only when solving Maxwell's equations in a particular set of boundary conditions.  If we can find a way to directly probe the local $\Jeff$, we will be insensitive to the global boundary conditions and any resonances they introduce.  To do so, let us revisit the argument of \secref{robustness}.  Since $m_\DM R>1$ now, the electric field in the radial direction will be unsuppressed.  However, so long as we have $m_\DM h\ll1$, the tangential electric field will still be suppressed globally.%
\footnote{\label{ftnt:Eparallel}%
Note that the upper end of the frequency range of interest has $m_\DM h\sim1$.  For these frequencies, the tangential electric field may not be suppressed everywhere in the Earth-ionosphere cavity.  However, if our measurements are taken on the ground, the tangential electric field will still be suppressed by virtue of the good (but imperfect) conductivity of the Earth's crust.  An incoming wave incident on an imperfect conductor generically yields a tangential electric field at the surface of the conductor which is suppressed by $\mathcal O(\sqrt{\omega/\sigma})$ [see Problem 7.4 of \citeR{Jackson}].  In the case of the Earth's crust, this suppression is $\lesssim2\%$ (for $f_\DM\lesssim1\,\mathrm{kHz}$ and $\sigma\gtrsim10^{-4}$\,S/m~\cite{atlas}).}
This means that if we take the tangential component of the Amp\`ere-Maxwell law \eqref{ampere}, then the second term can still be neglected, giving
\begin{equation}
    (\nabla\times\bm B)_\parallel=\bm J_{\mathrm{eff},\parallel}.
    \label{eq:parallel_ampere}
\end{equation}
We thus find that measuring the component of $\nabla\times\bm B$ parallel to the Earth's surface gives us a direct measurement of the local DM current!

In particular, this strategy does not even require measurements of $\nabla\times\bm B$ at multiple locations.  A single measurement can already be sensitive to ultralight DM. 
 Measurements of $\nabla\times\bm B$ at multiple locations across the Earth can still be useful though, since the DM should be phase-coherent over the entire Earth (so long as $\lambda_\mathrm{dB}\gg R$).

Naively, one may expect that it should be significantly more difficult to measure $\nabla\times\bm B$ than $\bm B$.  In particular, as we want to measure the parallel components of $\nabla\times\bm B$, we require measurements of the gradient of $\bm B$ \emph{in the vertical direction}.  Vertical separations of $d\sim100\,\mathrm{m}$ can be achieved by taking measurements at various altitudes along a hill/mountain.  However, since $\bm B$ only varies on lengthscales of roughly its Compton wavelength, $\lambda_\DM\lesssim1000\,\mathrm{km}$, we naively suffer a suppression of $\mathcal O(d/\lambda_\DM)\gtrsim10^{-4}$ compared to the low-frequency method of measuring only $\bm B$.  We note that this suppression also applies to physical magnetic-field noise sources though.  As the lower atmosphere does not efficiently conduct physical currents, we expect SM magnetic field sources in the lower atmosphere to have $(\nabla\times \bm B)_\parallel=0$.  Thus while the scheme of measuring the curl of $\bm B$ suppresses the signal relative to measuring $\bm B$ alone, it also acts as a noise rejection scheme, which should cancel external magnetic field noise sources.  The signal-to-noise ratio (SNR) relative to external correlated noise sources will therefore not suffer any suppression.  As we will see, if the cancellation scheme is implemented correctly, we expect the dominant noise source to then be internal magnetometer noise, which is uncorrelated between measurements.  (Note that the SNR relative to internal noise will still suffer the $\mathcal O(d/\lambda_\DM)$ suppression.)

\section{Measurement}
\label{sec:measurement}

In this section, we outline how to measure $(\nabla\times\bm B)_\parallel$.  In particular, we define an observable constructed from three nearby measurements of $\bm B$, which reproduces the curl of $\bm B$ (and is therefore sensitive to $\Jeff$).  We then evaluate this quantity for both the ultralight DM signal and for uncorrelated magnetometer noise.  As explained in \secref{highf}, if the cancellation scheme is implemented effectively, this should constitute the dominant noise source and ultimately set the sensitivity of our measurement scheme to ultralight DM. Figure~\ref{fig:sensitivity} shows the sensitivities to axion DM and DPDM, assuming the cancellation scheme has been implemented well enough that uncorrelated noise is indeed the dominant background.  In the remainder of the section, we consider various aspects of the implementation and enviromental properties, which will affect the efficiency of the cancellation scheme.  We outline benchmarks which must be met in order to achieve the sensitivities shown in \figref{sensitivity}.

\subsection{Construction of the curl}
\label{sec:curl}

Let us begin by defining our estimate of $(\nabla\times\bm B)_\parallel$.  As we are ultimately attempting to probe $\bm J_{\mathrm{eff},\parallel}$, we are only interested in the components of $\nabla\times\bm B$ which point along the projection of $\Jeff$ onto the Earth's surface.  In the DPDM case, $\Jeff$ can point in any direction, so both parallel directions are interesting.  In the axion case, however, \eqref{ax_current} indicates that $\Jeff$ can only point along $\bm B_\oplus$, and so we will only be interested in the north%
\footnote{Since the direction of $\Jeff$ is set by $\bm B_\oplus$, ``north" here should be understood as \emph{magnetic} north, which in general differs from geographic north.  Throughout the remainder of this work, we will always use ``north" and ``east" to refer to magnetic north/east.}
component of $\nabla\times\bm B$.  For the remainder of this subsection, we therefore focus on how to measure the north component of $\nabla\times\bm B$.  (Adapting the scheme to measure the east component of $\nabla\times\bm B$ is straightforward.)

\begin{figure*}[t]
\includegraphics[width=\textwidth]{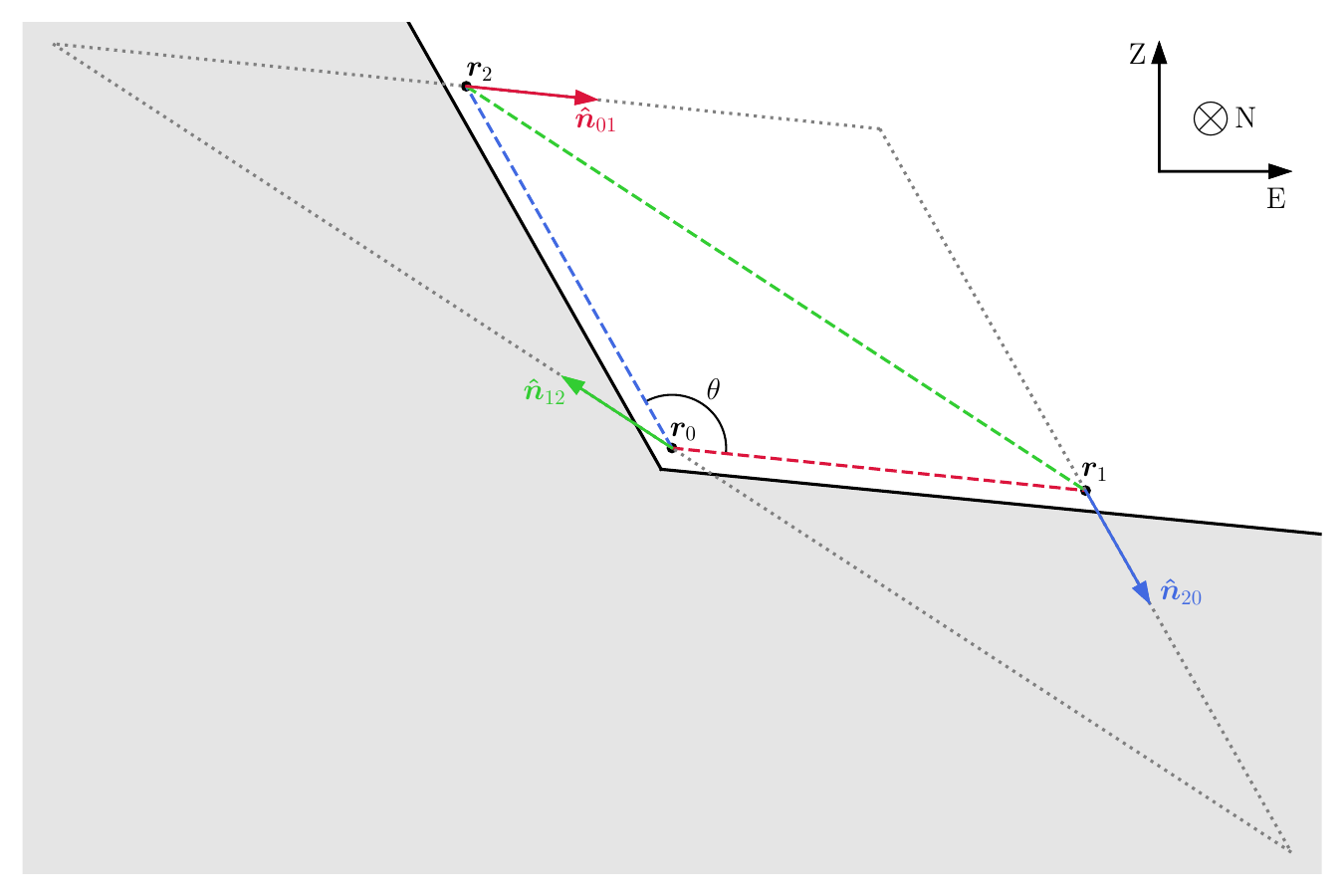}
\caption{\label{fig:curlfig}%
    Curl measurement scheme.  The scheme requires three single-axis magnetic field measurements at locations $\bm r_0,\bm r_1,\bm r_2$, lying in the vertical-east plane (in order to measure $\nabla\times\bm B$ in the north direction).  Such a configuration could, for instance, be achieved near a hill/mountain.  The direction of each magnetic-field measurement is given by the baseline between the other two locations.  The unit vectors $\nhat_{12},\nhat_{20},\nhat_{01}$ indicating the measurement directions are color-coded to match their corresponding baselines.  This scheme can be understood as a discretized line integral of $\bm B$ around the large dotted grey triangle.
    }
\end{figure*}

An estimate of $\nabla\times\bm B$ in the north direction requires measurements of the gradient of $\bm B$ in both the vertical and east directions.  These can be acquired with three single-axis magnetic field measurements at different locations/directions in the vertical-east plane.  Crucially, since gradients in the vertical direction are required, at least two of these locations must differ in altitude, meaning an ideal location would be near a hill/mountain.  Figure~\ref{fig:curlfig} shows an example configuration of three locations
\begin{align}
\bm r_0&=(x_0,y_0,z_0)\\
\bm r_1&=(x_1,y_0,z_1)\\
\bm r_2&=(x_2,y_0,z_2)
\end{align}
arranged in the vertical-east plane.  Here $x$ corresponds to the east direction (geomagnetic longitude), $y$ corresponds to the north direction (geomagnetic latitude), and $z$ corresponds to the vertical direction (altitude).  Note that all locations have the same $y$-coordinate.

The relevant component of the magnetic field at each location is given by the direction of the baseline formed by the other two locations (see \figref{curlfig}).  That is, at $\bm r_0$, we should measure the magnetic field in the direction
\begin{equation}
    \nhat_{12}=\frac{\bm r_2-\bm r_1}{|\bm r_2-\bm r_1|}.
\end{equation}
Likewise, at $\bm r_1$ and $\bm r_2$, we should measure the magnetic fields in the similarly defined $\nhat_{20}$ and $\nhat_{01}$ directions.

From these three measurements, we define the quantity
\begin{align}
    \Delta&=\bm B(\bm r_0)\cdot(\bm r_2-\bm r_1)+\bm B(\bm r_1)\cdot(\bm r_0-\bm r_2)\nl
    +\bm B(\bm r_2)\cdot(\bm r_1-\bm r_0).
    \label{eq:Delta_def}
\end{align}
It is straightforward to show that in the limit $\bm r_1,\bm r_2\rightarrow\bm r_0$,
\begin{equation}
    \Delta=\left[(\bm r_2-\bm r_0)\times(\bm r_1-\bm r_0)\right]\cdot(\nabla\times\bm B)
    \label{eq:Delta}
\end{equation}
(see \appref{leading} for derivation).  Since the cross product in \eqref{Delta} points in the north direction, then $\Delta$ measures the north component of $\nabla\times\bm B$.  Note that this cross product measures (half) the area of the large dotted grey triangle shown in \figref{curlfig}.  This can be understood by interpretting our measurement scheme as a discretized line integral of $\bm B$ around this triangle.  The quantity $\Delta$ should then correspond to the current flux through this triangle (i.e., $\Jeff$ times the area of the triangle).  The cross product also indicates that the signal is enhanced when the directions $\nhat_{20}$ and $\nhat_{01}$ are close to perpendicular (for fixed distances $|\bm r_1-\bm r_0|,|\bm r_2-\bm r_0|$).

\subsection{Signal and magnetometer noise}
\label{sec:signal}

Now we consider how an ultralight DM signal would appear in the quantity $\Delta$.  In the axion case, we can find this by simply plugging in the axion effective current \eqref{ax_current} for $\nabla\times\bm B$ in \eqref{Delta}.  This yields
\begin{align}
    \label{eq:ax_Delta}
    &\Delta_a=-g_{a\gamma}m_aB_{\oplus,\parallel}\cdot d_{10}d_{20}\sin\theta\cdot\IM\left[a_0e^{-im_at}\right]\\
    &\quad\sim10^{-14}\,\mathrm{T\,m}\left(\frac{g_{a\gamma}}{10^{-8}\,\mathrm{GeV}^{-1}}\right)\left(\frac{B_{\oplus,\parallel}}{20\,\mu\mathrm T}\right)\left(\frac d{100\,\mathrm m}\right)^2,
    \label{eq:ax_Delta_est}
\end{align}
where $B_{\oplus,\parallel}$ is the north component of the local geomagnetic field, $d_{mn}=|\bm r_m-\bm r_n|$, and $\theta$ is the (positive) angle between $\nhat_{01}$ and $\nhat_{02}=-\nhat_{20}$.  The estimate in \eqref{ax_Delta_est} [and all subsequent estimates below] assumes $d=d_{10}=d_{20}$ and $\theta=135^\circ$.  Recall that, as with \eqref{ax_wave}, \eqref{ax_Delta} only applies for $t<T_\mathrm{coh}$.  On longer timescales, the phase and amplitude of the axion DM will drift.

In the DPDM case, $\Jeff$ can point in any direction.  A three-magnetometer scheme, as in \figref{curlfig}, can only be sensitive to one parallel component of $\Jeff$.  Additional magnetometers can be added to probe the second parallel direction.  However, note that this technique cannot probe $J_{\mathrm{eff},z}$ because $E_z$ is generically nonzero and so contributes to the perpendicular component of the Amp\`ere-Maxwell law \eqref{ampere}.  Here we will only consider the DPDM signal in a three-magnetometer scheme, oriented to probe the north component of $\Jeff$ (since this is the scheme which would be used to measure the axion signal).  In such a scheme, the DPDM signal would be
\begin{align}
    \label{eq:DPDM_Delta}
    &\Delta_{A'}=-\varepsilon m_{A'}^2\cdot d_{10}d_{20}\sin\theta\cdot\RE[A'_{0,y}e^{-im_{A'}t}]\\
    &\quad\sim2\times10^{-14}\,\mathrm{T\,m}\left(\frac\varepsilon{10^{-7}}\right)\left(\frac{m_{A'}}{10^{-12}\,\mathrm{eV}}\right)\left(\frac d{100\,\mathrm m}\right)^2,
\end{align}
(again only for $t<T_\mathrm{coh}$), where $A'_{0,y}$ is the north component of the DPDM amplitude.  In terms of the amplitudes appearing in \eqref{DPDM_wave}, it is given by
\begin{equation}
    A'_{0,y}=-\frac i{\sqrt2}(A'_{0,+}+A'_{0,-})
\end{equation}
(when the $z$-axis is taken to be the local vertical direction).

As noted at the end of \secref{highf}, we expect our measurement scheme to cancel most correlated external noise sources between the three magnetometers.  Thus assuming our scheme is implemented effectively, we expect the dominant noise source to be internal magnetometer noise, which should be uncorrelated between the three magnetometers.  Denoting the instrumental noise power spectral density (PSD) of the magnetometer at location $\bm r_i$ by $S_{B,i}$, we can see from \eqref{Delta_def}, that the magnetomer noise contribution to the noise PSD for $\Delta$ will be
\begin{align}
    \label{eq:mag_noise}
    &S_{\Delta,\mathrm{mag}}=S_{B,1}d_{21}^2+S_{B,2}d_{20}^2+S_{B,3}d_{10}^2,\\
    &\quad\sim5\times10^{-24}\,\mathrm{T}^2\,\mathrm{m^2/Hz}\left(\frac{S_{B,i}}{100\,\mathrm{fT}^2/\mathrm{Hz}}\right)\left(\frac d{100\,\mathrm m}\right)^2.
    \label{eq:mag_noise_est}
\end{align}

In the case that the total integration time of the measurement is less than the coherence time of the DM, $T_\mathrm{int}<T_\mathrm{coh}$, then \eqref[s]{ax_Delta} and (\ref{eq:DPDM_Delta}) apply as written.  In this case, the SNR can be computed by comparing $\Delta_\DM^2$ from these expressions directly to $S_{\Delta,\mathrm{mag}}$ in \eqref{mag_noise} divided by $T_\mathrm{int}$.  This will yield a sensitivity that scales like $g_{a\gamma},\varepsilon\sim1/\sqrt{T_\mathrm{int}}$.  In our frequency range of interest, however, $T_\mathrm{coh}$ is not very long, e.g. $T_\mathrm{coh}\sim1\,\mathrm{day}$ for $f_\DM=10\Hz$.  We thus expect to be able to integrate for longer than a coherence time.  In this case, we can treat each coherence time as an independent observation (for which \eqref[s]{ax_Delta} and (\ref{eq:DPDM_Delta}) are valid).  The SNR for the full $T_\mathrm{int}$ is then simply the SNR for each individual $T_\mathrm{coh}$ summed in quadrature over the independent observations
\begin{equation}
    \mathrm{SNR}=\frac{\Delta_\DM^2}{S_{\Delta,\mathrm{mag}}/T_\mathrm{coh}}\cdot\sqrt{\frac{T_\mathrm{int}}{T_\mathrm{coh}}}.
    \label{eq:SNR}
\end{equation}
In this case, the sensitivity will instead scale like $g_{a\gamma},\varepsilon\sim1/\sqrt[4]{T_\mathrm{int}T_\mathrm{coh}}$.  (See Appendix A\,5 of \citeR{budker2014proposal} for a similar discussion of the sensitivity scaling as a function of $T_\mathrm{int}$.)

Sensitivity to the ultralight DM signal can be enhanced by taking advantage of the spatial coherence of $\Jeff$.  Since $\lambda_\mathrm{dB}\gg R$, then any two simultaneous measurements of $\Jeff$ on Earth should be phase coherent.  In general if we implement the curl measurement scheme at $N$ distinct locations, this should enhance the SNR by a factor of $N$ (and so improve the sensitivity to $g_{a\gamma},\varepsilon$ by a factor of $\sqrt N$).

\begin{figure*}[t]
\includegraphics[width=0.49\textwidth]{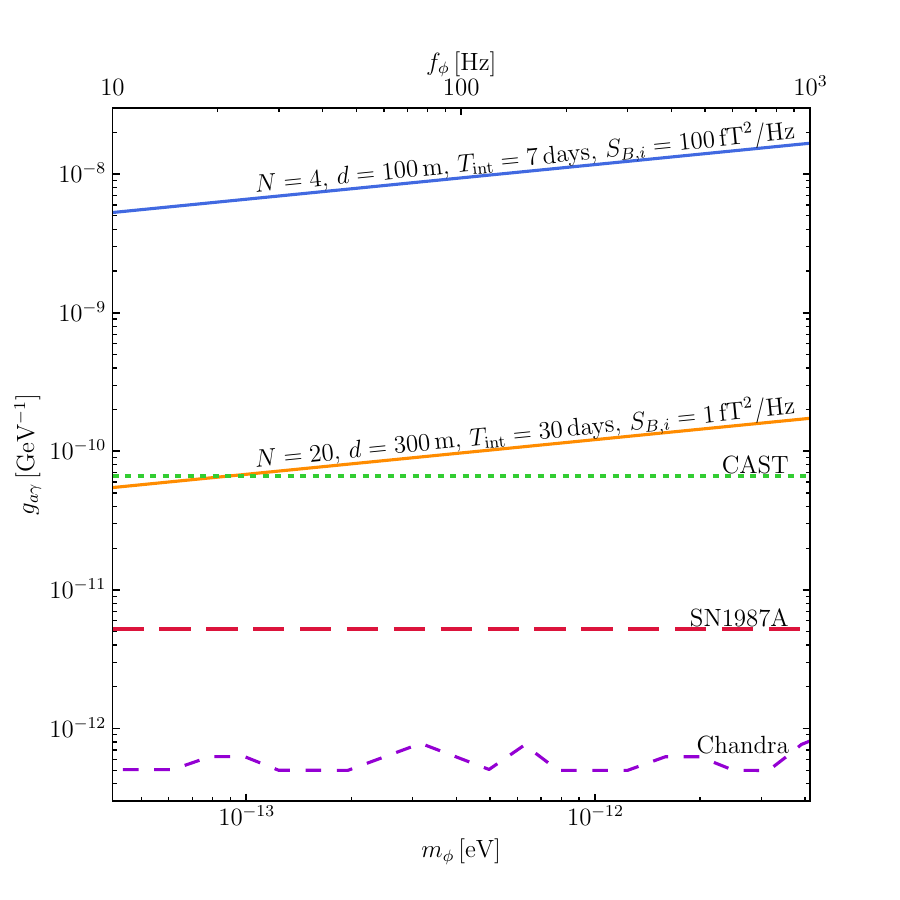}
\includegraphics[width=0.49\textwidth]{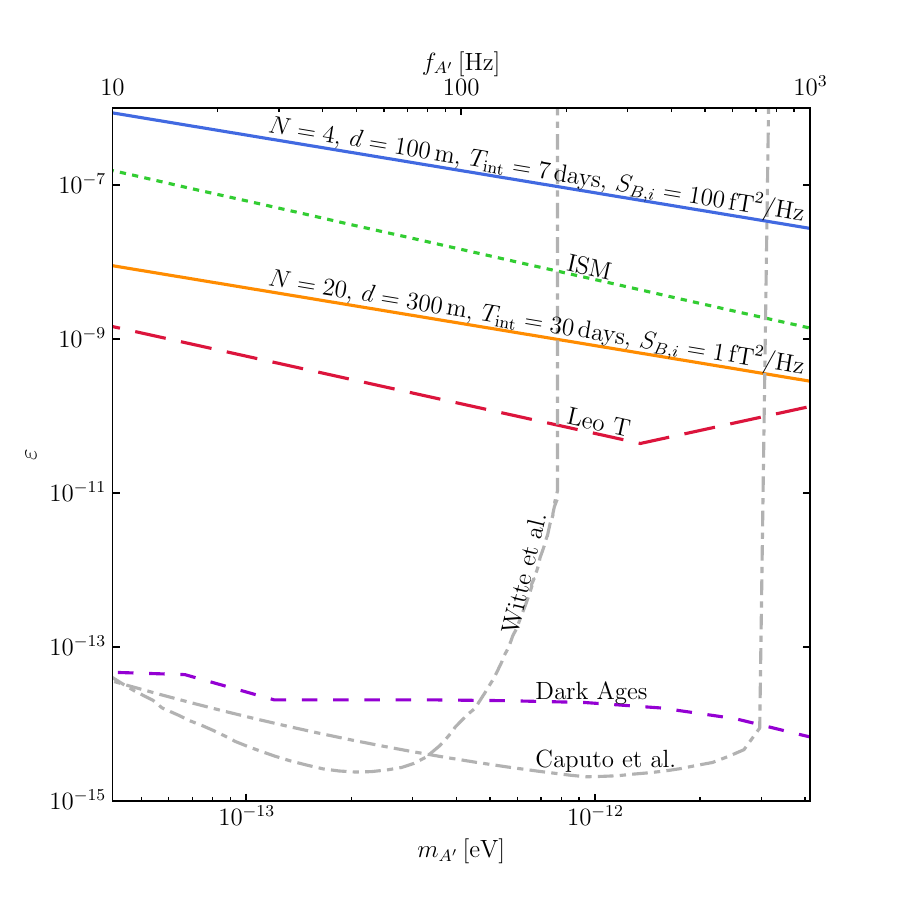}
\caption[]{\label{fig:sensitivity}%
    Projected sensitivity of curl measurement to axion DM (left) and DPDM (right), assuming that the cancellation scheme is implemented effectively enough that internal magnetometer noise is the dominant noise source.  Two projections (solid lines) are shown with different choices for: the number of independent curl measurements, $N$; the distance between magnetometers (within one measurement scheme), $d$; the integration time, $T_\mathrm{int}$; and the internal magnetometer noise PSD, $S_{B,i}$.  One projection corresponds to representative parameters for the upcoming 2024 SNIPE Hunt run (blue), and the other corresponds to more optimistic parameters for future runs (orange).  In addition to these, we take the parallel geomagnetic field $B_{\oplus,\parallel}=20\,\mu\mathrm{T}$, the angle between the baselines $\theta=135^\circ$, and a threshold $\mathrm{SNR}=3$ in these estimates.  We also show existing constraints as dashed curves.\textsuperscript{a} The axion constraints include limits from: the CAST helioscope search for solar axions (green)~\cite{Anastassopoulos:2017ftl}; non-observation of gamma rays in coincidence with SN1987A (red)~\cite{Hoof_2023}; and X-ray observations of the quasar H1821+643 from the Chandra telescope (purple)~\cite{10.1093/mnras/stab3464}.  The DPDM constraints include limits from: heating of the interstellar medium (green)~\cite{Dubovsky:2015cca}; heating of the dwarf galaxy Leo T (red)~\cite{Wadekar:2019xnf}; resonant conversion of DPDM during the dark ages (purple)~\cite{McDermott:2019lch}, and resonant conversion during the epoch of helium reionization (grey).\textsuperscript{b} \citeR[s]{Witte:2020rvb,Caputo:2020bdy} find different limits for this final effect, so we include both for completeness.\textsuperscript{c} Note that the CAST bound is the only existing laboratory-based constraint shown.  The rest are all astrophysical/cosmological in nature.  Our proposal therefore offers a complementary probe in these regions of ultralight DM parameter space.
    }
    \raggedright\footnotesize\textsuperscript{a} Several of these limits were acquired from \citeR{Oharegithub,Caputo:2021eaa}.\\
    \textsuperscript{b} We note two additional constraints which are not shown. \citeR{escudero2023axion} constrains axion DM based on reionization of the intergalactic medium by axion stars.  This constraint depends strongly on modeling of the axion star abundance.  There also exist constraints on both axion DM and DPDM in this mass range from black hole superradiance~\cite{Cardoso_2018}.  These constraints can be modified by self-interactions~\cite{Baryakhtar:2020gao} or interactions with the Standard Model~\cite{Siemonsen_2023}.\\
    \textsuperscript{c} The primary source of disagreement between these two limits is the manner in which the energy deposited by the resonant conversion redistributes itself.  \citeR{Witte:2020rvb} (Witte et al.) assumes that the energy is deposited locally, while \citeR{Caputo:2020bdy} (Caputo et al.) assumes that it is redistributed evenly across the universe.  As the issue of energy transport in these contexts is complex, we take no stance on this disagreement and present both limits.
\end{figure*}

Figure~\ref{fig:sensitivity} shows the potential reach of our measurement scheme, assuming that internal magnetometer noise dominates.  The sensitivities are computed by setting $\mathrm{SNR}=3$ in \eqref{SNR} [and including the enhancement factor of $N$ described above].  The left panel shows our projected sensitivity (solid lines) to axion DM, along with existing constraints (dashed lines).  The right panel shows the same for DPDM.  We show two projections in each plot.  The blue lines use parameters which are representative of the search that the SNIPE Hunt collaboration will undertake in 2024.  The orange lines show optimistic parameters which may be considered for future runs.  The axion sensitivities become stronger at lower frequencies because of the longer coherence time of the axion signal.  Meanwhile, the DPDM sensitivies become weaker because of the $m_{A'}$ dependence in \eqref{DPDM_Delta} [which outweighs the coherence time enhancement].  We note that all the existing constraints shown in \figref{sensitivity}, except for the CAST bound, are astrophysical/cosmological in nature.  Our scheme therefore offers a new way to probe ultralight DM parameter space with terrestrial measurements.

\subsection{Finite difference noise}
\label{sec:finite_difference}

The projected sensitivities shown in \figref{sensitivity} assume that the dominant noise source for our measurement of $\Delta$ is internal magnetometer noise.  As mentioned at the end of \secref{highf}, this should be the case if the curl measurement scheme is implemented effectively, as we expect $(\nabla\times \bm B)_\parallel=0$ for enviromental magnetic field sources. In the remainder of this section, we outline the implementation benchmarks and enviromental conditions which must be met in order for other contributions to $\Delta$ to be subdominant to magnetometer noise.

We begin with the issue of uncanceled enviromental magnetic field noise.  This can arise due to the fact that \eqref{Delta_def} is a discretized estimate of the curl, and only approaches \eqref{Delta} in the limit where the magnetometers are close to one another.  Therefore even if all environmental noise actually satisfies $(\nabla\times\bm B)_\parallel=0$, our measurement of $\Delta$ can be nonzero due to the finite distance between our magnetometers.  We refer to this as finite difference noise.

As shown in \appref{nlo}, there are second-order corrections to \eqref{Delta} [which do not generically cancel].  As can be seen in \eqref{second_order}, these contributions generically have the form $\partial^2B\cdot d^3$, where $d$ is the characteristic separation between magnetometers.  Enviromental magnetic fields of frequency $f$ should vary on lengthscales of roughly their wavelength $\lambda=1/f$, and so $\partial^2B\sim B/\lambda^2$.  Then if the enviromental noise PSD in the vicinity of the measurement location is $S_{B,\mathrm{ext}}$, the finite difference noise contribution to $\Delta$ will be roughly
\begin{align}
    \label{eq:FDnoise}
    &S_{\Delta,\mathrm{FD}}\sim\frac{S_{B,\mathrm{ext}}d^6}{\lambda^4}\\
    &\quad\sim10^{-28}\,\mathrm{T}^2\,\mathrm{m^2/Hz}\left(\frac{S_{B,\mathrm{ext}}}{1\,\mathrm{nT}^2/\mathrm{Hz}}\right)\left(\frac d{100\,\mathrm m}\right)^6\left(\frac f{1\,\mathrm{kHz}}\right)^4.
    \label{eq:FDnoise_est}
\end{align}
We can see that finite difference noise will be significantly smaller than the magnetometer noise estimated in \eqref{mag_noise_est}, so long as the measurement is performed in a sufficiently radio-quiet location.  The previous SNIPE Hunt run performed in summer 2022 already observed enviromental noise levels $\lesssim\mathrm{nT}^2/\mathrm{Hz}$~\cite{sulai2023hunt}, so we anticipate this to be an achievable requirement.

\subsection{Directional uncertainty}
\label{sec:direction}

A key implementation requirement will be precision in the directions of the magnetometer measurements.  The desired orientations of the magnetometers are shown in \figref{curlfig}, where the direction of each measurement is set by the baseline between the other two magnetometers.  Errors in the orientation of the magnetometers will disrupt the cancellation that yields \eqref{Delta} from \eqref{Delta_def}.  Then if there are local enviromental magnetic fields $\bm B_\mathrm{ext}$, the quantity $\Delta$ will receive contributions directly from $\bm B_\mathrm{ext}$ [even if $(\nabla\times\bm B_\mathrm{ext})_\parallel$ vanishes].  If the magnetometers are misaligned by an angle of $\epsilon$, this uncanceled noise will generically be $\Delta_\mathrm{dir}\sim\epsilon B_\mathrm{ext}d$ (see \appref{direction}).  Quantitatively, this will yield a noise level of
\begin{align}
    S_{\Delta,\mathrm{dir}}&\sim\,3\times10^{-24}\,\mathrm{T}^2\,\mathrm{m^2/Hz}\left(\frac\epsilon{1^\circ}\right)^2\nl
    \cdot\left(\frac{S_{B,\mathrm{ext}}}{1\,\mathrm{pT}^2/\mathrm{Hz}}\right)\left(\frac d{100\,\mathrm m}\right)^2.
\end{align}
Note that here, we require a lower $S_{B,\mathrm{ext}}$ than in \eqref{FDnoise_est}, in order for this noise source to be subdominant to \eqref{mag_noise_est}.  Preliminary measurements show that such low noise levels are achievable in sufficiently quiet locations.

If the local environment is sufficiently radio quiet, then orienting one single-axis magnetometer at each location with degree-level accuracy may suffice to make this uncanceled noise subdominant to internal noise.  (Over a baseline of $d\sim100\,\mathrm{m}$, angular precision of $\epsilon\sim1^\circ$ requires precision in position of $\sim1\,\mathrm{m}$, which can be achieved using GPS~\cite{gps}.)  If however, the local magnetic field noise is high enough that better directional precision is required, it may be more advantageous to utilize multiple single-axis magnetometers (or one multi-axis magnetometer) at each location.  Then the orientation of the magnetometers can be treated as a free parameter in the data analysis to account for misalignment of the magnetometers.  This will then allow for full cancellation of environmental noise.

\subsection{Timing uncertainty}
\label{sec:timing}

Another key aspect of the implementation will be accurately identifying the time at which each magnetic field measurement is recorded.  Each measurement location must be equipped with a clock to set the measurement cadence, and errors in the stability of this clock may also disrupt the cancellation that yields \eqref{Delta}.  Specifically, if the timing uncertainty of each clock is $\delta t$, then errors in the timing will lead to relative phase offsets between the stations of $\sim\omega\delta t$ (where $\omega=2\pi f$), and so ultimately a noise level of
\begin{align}
    \label{eq:time_unc}
    S_{\Delta,\mathrm{time}}&\sim S_{B,\mathrm{ext}}d^2\cdot(\omega\delta t)^2\\    &\sim4\times10^{-25}\,\mathrm{T}^2\,\mathrm{m^2/Hz}\left(\frac{S_{B,\mathrm{ext}}}{100\,\mathrm{pT}^2/\mathrm{Hz}}\right)\nl
    \cdot\left(\frac d{100\,\mathrm m}\right)^2\left(\frac f{1\,\mathrm{kHz}}\right)^2\left(\frac{\delta t}{100\,\mathrm{ns}}\right)^2.
    \label{eq:time_unc_est}
\end{align}
As with directional uncertainty, the uncanceled noise from timing uncertainty will depend heavily on the ambient magnetic field noise.  Standard data acquistion systems can achieve clock stabilities of $\sim100$\,ns~\cite{Wlo14daqTiming,NI_DAQ_GPS_2023}, so we expect that timing uncertainty should be less of an issue than directional uncertainty.  Should better timing precision be required, we note that atomic clocks can achieve sub-nanosecond stability~\cite{FS725}.

In addition to timing uncertainty from the stability of each individual clock, there exists a distinct issue of how well the clocks at different locations can be synchronized at the beginning of the experiment.  Using GPS, this can be achieved with an uncertainty of about $30\,\mathrm{ns}$~\cite{gps}, so we anticipate that synchronizing clocks should be no more difficult than maintaining the stability of each individual clock.  If synchronization does however prove to be an issue, we note that this offset in timing between stations should not drift (by more than $\delta t$) over the course of the experiment, and so it can be accounted for by introducing additional parameters that characterize the offset into the analysis.  This should allow the relative timing between stations to be adjusted to achieve full noise cancellation.  It is also worth noting that the timing uncertainty noise in \eqref{time_unc} is frequency-dependent, so even if timing uncertainty does prove to be a dominant noise source, it should only be an issue at higher frequencies.

Finally, we also note that details of the magnetometers and electronics can lead to frequency-dependent phase shifts (e.g., due to high/low-pass filtering) which may differ between magnetometer setups.  We anticipate such effects may introduce up to $\sim\mathrm{mrad}$ relative phase offsets between different locations, which would translate to noise levels of $\sim10^{-24}\,\mathrm{T}^2\,\mathrm{m^2/Hz}$, for $S_{B,\mathrm{ext}}$ and $d$ as in \eqref{time_unc_est}.  The effect of such phase offsets will be difficult to remove during data analysis, as they can depend on frequency and can drift in time along with environmental factors such as temperature.  We leave further exploration of this issue to experimental implementations of this measurement scheme.

\subsection{Vibrational noise}
\label{sec:vibrations}

We anticipate that the most difficult implementation requirement will likely be vibration isolation.  Small vibrations due to seismic noise or human activity can cause rotations of the magnetometers relative to the Earth's local geomagnetic field $B_\oplus\sim50\,\mu\mathrm{T}$.  These rotations will then lead to variations in the measured magnetic field.  Specifically, if the magnetometer experiences angular rotational noise (pointing noise) $S_{\delta\theta}$ due to vibrations, then the resulting noise contribution to $\Delta$ will be
\begin{align}
    &S_{\Delta,\mathrm{vib}}\sim S_{\delta\theta}B_\oplus^2d^2\\
    &\quad\sim3\times10^{-24}\,\mathrm{T}^2\,\mathrm{m^2/Hz}\left(\frac{S_{\delta\theta}}{10^{-19}\,\mathrm{rad}^2/\mathrm{Hz}}\right)\left(\frac d{100\,\mathrm m}\right).
    \label{eq:vib_noise_est}
\end{align}
For a magnetometer of roughly $1\,\mathrm m$, the above angular rotational noise corresponds to a vibrational noise of $0.1\,\mathrm{nm}^2/\mathrm{Hz}$.  It is clear that very good vibration isolation will be required in order for vibrational noise to be subdominant to the internal magnetometer noise in \eqref{mag_noise_est}.

Luckily, seismic noise is relatively suppressed in much of our frequency range of interest.  For instance, one of the dominant noise sources for ground vibrations at high frequencies is wind.  Typical wind speeds lead to vibrational noise of $\lesssim1\,\mathrm{nm}^2/\mathrm{Hz}$ for frequencies $\gtrsim10\,\mathrm{Hz}$~\cite{naderyan,withers}.  (The direct effect of wind on the magnetometer orientation can be shielded by appropriately covering the magnetometer.)  We thus anticipate that natural high-frequency sources of vibrations may not require much damping.  Lower-frequency or man-made sources may, however, dominate the vibrational noise.

Experiments in the field have been able to demonstrate that with careful choice of location and stable installation, induction coil magnetometers can reach a noise floor of $\sim 10$~fT/$\sqrt{{\textrm{Hz}}}$ \cite{LEMI_manual_2020}, see also, for example, Ref.~\cite{coughlin2018measurement}. This shows that, in principle, vibrational noise can be controlled at roughly the level of the proposed intrinsic magnetometer sensitivity given by \eqref{mag_noise_est}. Furthermore, telescopes with careful mounting, damping, and isolation show vibrational noise in the range of milliarcseconds (corresponding to $\sim 5\times 10^{-9}$~rad/$\sqrt{{\textrm{Hz}}}$) in the frequency range of interest for the experiment \cite{Alt01telescopeVibrations}. This shows that reducing the vibrational noise to the required level is, while challenging, within reach.

\subsection{Local electric field direction}
\label{sec:efields}

A key assumption underlying the noise cancellation of our technique is that the local electric field is exactly vertical.  Namely, we have claimed that $E_y=0$ and that environmental noise sources should have $\bm J=0$ (see next subsection for a discussion of this assumption), and so therefore
\begin{equation}
    (\nabla\times\bm B)_y=J_y-\partial_tE_y=0
\end{equation}
for environmental noise.  The claim that $E_y=0$ relies on the perfectly spherical model of the Earth-ionosphere cavity, which enforces long-wavelength electric fields to point exactly radially.  (For higher frequencies, the electric field may not be vertical everywhere, but should still be vertical near the surface of the Earth, where our measurement is being performed; see footnote~\ref{ftnt:Eparallel}.)

As noted in \secref{robustness}, the surfaces bounding this cavity may not really be perfectly spherical, and thus the actual direction of the electric fields associated with environmental noise sources may not exactly match the local vertical direction.  If the local electric field direction deviates from the local vertical by $\epsilon$, then we instead have $E_y+\epsilon E_z=0$, and so likewise
\begin{equation}
    (\nabla\times\bm B)_y=-\epsilon(\nabla\times\bm B)_z\sim\epsilon B/\lambda.
\end{equation}
This leads to contributions to $\Delta$ of roughly
\begin{align}
    S_{\Delta,E\text{-dir}}&\sim\frac{\epsilon^2S_{B,\mathrm{ext}}d^4}{\lambda^2}
    \\&\sim3\times10^{-25}\,\mathrm{T}^2\,\mathrm{m^2/Hz}\left(\frac\epsilon{1^\circ}\right)^2\left(\frac{S_{B,\mathrm{ext}}}{1\,\mathrm{nT}^2/\mathrm{Hz}}\right)\nl
    \cdot\left(\frac d{100\,\mathrm m}\right)^4\left(\frac f{1\,\mathrm{kHz}}\right)^2.
    \label{eq:Edir_est}
\end{align}

Deviations in the local electric field direction may arise, for instance, from asphericities in the Earth's geometry.  Asphericities on lengthscales shorter than a wavelength $\lambda\gtrsim300\,\mathrm{km}$ (for $f=1\,\mathrm{kHz}$) should not significantly affect the electric field solution.  Therefore the effects of individual hills/mountains/clouds should be negligible.  Instead to estimate $\epsilon$, we should consider what elevation gains are possible along lengthscales of $\lambda$.%
\footnote{Another contribution to $\epsilon$ may come from the global oblateness of the Earth.  This oblateness is only about $0.3\%$~\cite{groten_2000}, so the estimate it would give for $\epsilon$ should be even smaller than the one in \eqref{vert_angle}.}
The largest such elevation gains should be at most $\sim\mathrm{few\,km}$.  Therefore the largest deviation of the local electric field direction from the local vertical should be roughly
\begin{equation}
    \epsilon\lesssim\frac{\mathrm{few\,km}}{300\,\mathrm{km}}\sim1^\circ.
    \label{eq:vert_angle}
\end{equation}
From the estimate in \eqref{Edir_est}, we see that this deviation is sufficiently small that its effect is still subdominant to magnetometer noise.

\subsection{Atmospheric currents}
\label{sec:currents}

Our noise cancellation scheme also relies on the assumption that there are no significant atmospheric currents at our frequencies of interest.  In fact, the lower atmosphere does maintain a small DC current due to its slight conductivity and the ever-present vertical electric field between the negatively charged Earth's surface and postively charged ionosphere.  During ``fair weather" conditions, the lower atmosphere has a conductivity of $\sim10^{-14}\,\mathrm{S/m}$ and electric field of $\sim100\,\mathrm{V/m}$, leading to a vertical DC current of $\sim\mathrm{pA/m^2}$~\cite{harrison,SIINGH200791}.  We note that this current is not directly relevant for our measurement, as we will only be senstive to currents which are parallel to the surface and which vary at frequencies $10\,\mathrm{Hz}\lesssim f\lesssim1\,\mathrm{kHz}$.  Currents of this kind in the lower atmosphere have not been particularly well studied.  We can, however, use the vertical DC current to derive a very crude upper limit on such currents, based on the requirement that parallel current fluctuations should not significantly exceed the vertical DC current.  This yields
\begin{align}
    S_{\Delta,\mathrm{currents}}&\sim S_{J_y}d^4\lesssim\frac{(\mathrm{pA/m^2})^2d^4}f\\
    &\sim10^{-29}\,\mathrm{T}^2\,\mathrm{m^2/Hz}\left(\frac d{100\,\mathrm m}\right)^4\left(\frac{10\,\mathrm{Hz}}f\right).
\end{align}
Therefore, the presence of lower atmospheric currents should not affect our measurement scheme, during fair weather conditions.  We do note that extreme weather conditions, e.g. lightning, can create large currents in the lower atmosphere which could lead to uncanceled noise.  These noise sources should be easily identifiable in the data, and noisy data corresponding to weather events can be excluded from our analysis.

\section{Conclusion}
\label{sec:conclusion}

Axions and dark photons are well-motivated candidates for DM over a wide range of possible masses.  For masses $m_\DM\lesssim10^{-11}\,\mathrm{eV}$, there are few existing laboratory probes of these DM candidates.  \citeR[s]{Fedderke:2021rys,Arza_2022} described one such probe, an oscillating magnetic field signal at the Earth's surface sourced by ultralight DM.  It was shown that a component of this global signal was robust to environmental details for $m_\DM\ll R^{-1}\sim3\times10^{-14}\,\mathrm{eV}$, and so constraints could be placed on ultralight DM using a global array of synchronized magnetometers.

In this work, we described a new measurement scheme which can allow this effect to be robustly probed at higher DM masses.  Although the global signal $\bm B$ depends on the boundary conditions of the near-Earth environment, the component of $\nabla\times\bm B$ parallel to the Earth's surface does not.  This is because the parallel electric field $\bm E_\parallel$ vanishes (to leading order), and so $(\nabla\times\bm B)_\parallel$ is directly related the dark matter effective current $\bm J_{\mathrm{eff},\parallel}$ by the Amp\`ere-Maxwell law \eqref{parallel_ampere}.  By taking multiple magnetic field measurements at nearby locations on a hill/mountain, as shown in \figref{curlfig}, we can measure this curl, and so directly probe the local DM amplitude.  In addition, because we expect environmental magnetic fields to have $\nabla\times\bm B=0$, our technique should act as a background rejection scheme for external environmental noise.

In \figref{sensitivity}, we show the projected sensitivities of our scheme to axion DM and DPDM, assuming that our sensitivity is dominated by internal magnetometer noise.  We show one projection using representative parameters for the upcoming 2024 SNIPE Hunt run, and one using optimistic future parameters.  In the DPDM case, our measurement scheme has the potential to be the strongest laboratory probe of DPDM in the $10\,\mathrm{Hz}\leq f_\DM\leq1\,\mathrm{kHz}$ frequency range.  In the axion DM case, our optimistic projection approaches sensitivities comparable to the leading laboratory contraints from the CAST helioscope.

Throughout \secref{measurement}, we outlined several benchmarks that should be met in order to reach the projections shown in \figref{sensitivity}.  Of them, the two most important will likely be precision in the orientation of the magnetometers and vibrational isolation.  For the former, degree-level precision will be required if the ambient environmental magnetic field noise is $\sim\mathrm{pT}^2/\mathrm{Hz}$.  (For comparison, the 2022 SNIPE Hunt experimental run observed noise levels $\lesssim\mathrm{nT}^2/\mathrm{Hz}$.)  If the magnetic field noise is larger, three-axis magnetometer measurements may be utilized in order to account for uncertainties in orientation.  For vibrational isolations, a vibrational noise level of $0.1\,\mathrm{nm}^2/\mathrm{Hz}$ will likely be required for a meter-long magnetometer.  We expect the ambient vibrational noise at these frequencies to be not too much larger than this, so minimal damping will be required.

The SNIPE Hunt Collaboration intends to begin implementing this technique in their 2024 experimental run, which will consist of temporary magnetometer setups obtaining a few independent curl measurements over the course of a few days.  The logistics of maintaining temporary magnetometers limits the baselines between magnetometers in this experiment.  In the further future, permanent setups in radio-quiet locations may allow for longer baselines, more independent curl measurements, and much longer integration times, thereby increasing the sensitivity of this technique.

Finally, in addition, to axion DM and DPDM, we note that ultralight millicharged DM also exhibits the Earth-transducer effect, as it too can be associated with an effective current.  In the spherical model used in \citeR[s]{Fedderke:2021rys,Arza_2022}, the global signal of millicharged DM exhibits an explicit dependence on the height of the atmosphere.  Since this height is a model-dependent quantity, the global millicharged DM signal cannot be robustly predicted, and as such, the searches in \citeR[s]{Arza_2022,Fedderke:2021iqw} cannot derive robust limits on millicharged DM.  However, as shown in this work, the curl of the magnetic field is a local measurement of the DM current, and so should not depend on the height of the atmosphere.  The technique outlined in this work is therefore more suitable for detecting millicharged DM.  We leave further exploration of this application to future work.

\acknowledgments

We thank Derek Jackson Kimball, Abaz Kryemadhi, Jason Stalnaker, and Ibrahim Sulai for consultation on experimental aspects of the scheme proposed in this work.  We also thank Ariel Arza, Michael Fedderke, Peter Graham, and Jedidiah Thompson for insights into the theoretical details of our scheme.

We thank the Patras workshop for facilitating the conception of this work, as well as the Aspen Center for Physics, which is supported by NSF Grant No.~PHY-2210452, for hospitality during this work's final stages.

S.K. is also supported by the U.S. Department of Energy, Office of Science, National Quantum Information Science Research Centers, Superconducting Quantum Materials and Systems Center (SQMS) under contract number DE-AC02-07CH11359.

I.B is thankful to the Atomic Physics Gordon Research Conference for its hospitality during the late stages of this work.

\appendix

\section{Contributions to $\Delta$}
\label{app:contributions}

In this appendix, we expand the definition of $\Delta$ given in \eqref{Delta_def} in the long-wavelength limit to determine the various contributions to $\Delta$.  First, we compute the leading order contribution to find \eqref{Delta}.  Then for sake of completenes, we expand the next-to-leading order contributions to demonstrate that no generic cancellation occurs.  Therefore the estimate used for the finite difference noise in \eqref{FDnoise} is appropriate.  Finally, we show how magnetometer misalignment affects the magnetic field cancellation, and compute the uncanceled noise contributions to $\Delta$.

\subsection{Leading order}
\label{app:leading}

Let us assume that $\bm B$ varies on lengthscales much larger than the relevant baselines, $\lambda\gg d_{10},d_{20}$.  Then we may Taylor expand $\bm B$ to leading order around $\bm r_0$
\begin{align}
    B_i(\bm r)&\approx B_i(\bm r_0)+\partial_xB_i(\bm r_0)(x-x_0)\nl
    +\partial_yB_i(\bm r_0)(y-y_0)+\partial_zB_i(\bm r_0)(z-z_0),
    \label{eq:leading}
\end{align}
for $i=x,y,z$.  If we plug this into \eqref{Delta_def}, the leading order terms will cancel.  The remaining first order contributions will then be
\begin{align}
    \Delta=\sum_{i,j=x,y,z}&\Bigg(\partial_jB_i(\bm r_0)(r_{1,j}-r_{0,j})(r_{0,i}-r_{2,i})\nl
    +\partial_jB_i(\bm r_0)(r_{2,j}-r_{0,j})(r_{1,i}-r_{0,i})\Bigg),
\end{align}
where $r_{n,i}$ represents the $i$-th component of $\bm r_n$.  Clearly the $i=j$ terms in this sum vanish.  Recall that all three locations have the same $y$-coordinate, so the $i=y$ and $j=y$ terms also vanish.  Finally, we are left with
\begin{align}
    \Delta&=-\partial_xB_z(\bm r_0)(x_1-x_0)(z_2-z_0)\nl
    +\partial_xB_z(\bm r_0)(x_2-x_0)(z_1-z_0)\nl
    -\partial_zB_x(\bm r_0)(x_2-x_0)(z_1-z_0)\nl
    +\partial_zB_x(\bm r_0)(x_1-x_0)(z_2-z_0)\\
    &=\left[\nabla\times\bm B(\bm r_0)\right]_y\left[(\bm r_2-\bm r_0)\times(\bm r_1-\bm r_0)\right]_y,
\end{align}
since $(\bm r_2-\bm r_0)\times(\bm r_1-\bm r_0)$ points in the $\bm y$-direction, then this is equivalent to \eqref{Delta}.

\subsection{Next-to-leading order}
\label{app:nlo}

The second order corrections to \eqref{leading} are
\begin{equation}
    B_i^{(2)}(\bm r_0)=\frac12\sum_{j,k}\partial_j\partial_kB_i(\bm r_0)(r_j-r_{0,j})(r_k-r_{0,k}).
\end{equation}
Then the corresponding corrections to $\Delta$ are
\begin{widetext}
\begin{align}
    \Delta^{(2)}&=\frac12\sum_{i,j,k}\left(\partial_j\partial_kB_i(\bm r_0)(r_{1,j}-r_{0,j})(r_{1,k}-r_{0,k})(r_{0,i}-r_{2,i})+\partial_j\partial_kB_i(\bm r_0)(r_{2,j}-r_{0,j})(r_{2,k}-r_{0,k})(r_{1,i}-r_{0,i})\right)\\
    &=-\frac12\partial_x^2B_x(\bm r_0)(x_1-x_0)(x_0-x_2)(x_2-x_1)-
    \frac12\partial_z^2B_z(\bm r_0)(z_1-z_0)(z_0-z_2)(z_2-z_1)\nl
    +\frac12\partial_x^2B_z(\bm r_0)\left[(x_1-x_0)^2(z_0-z_2)+(x_2-x_0)^2(z_1-z_0)\right]\nl
    +\frac12\partial_z^2B_x(\bm r_0)\left[(z_1-z_0)^2(x_0-x_2)+(z_2-z_0)^2(x_1-x_0)\right]\nl
    -\partial_x\partial_zB_x(\bm r_0)(x_1-x_0)(x_0-x_2)(z_2-z_1)-\partial_x\partial_zB_z(\bm r_0)(z_1-z_0)(z_0-z_2)(x_2-x_1)
    \label{eq:second_order}
\end{align}
\end{widetext}
Note that generically no cancellation occurs between these terms.  Therefore, the second order corrections indeed dominate the finite difference noise.  If $\bm B$ varies on lengthscales of roughly the wavelength $\lambda$, then $\partial^2B\sim B/\lambda^2$.  If, moreover, the characteristic separation between stations is roughly $d$, then all terms in \eqref{second_order} are $\mathcal O(Bd^3/\lambda^2)$, as estimated in \secref{finite_difference}.

\subsection{Directional uncertainties}
\label{app:direction}

If the magnetometers are oriented incorrectly, the calculation of \appref{leading} changes.  Suppose the magnetometer at location $\bm r_n$ is misaligned by an angle $\epsilon_n$ in the $xz$-plane, and by an angle $\delta_n$ in the $y$-direction.  Then $\Delta$ in \eqref{Delta_def} would be calculated as
\begin{align}
    \Delta&=d_{21}\bm B(\bm r_0)\cdot(\nhat_{12}+\epsilon_0\Y\times\nhat_{12}+\delta_0\Y)\nl
    +d_{20}\bm B(\bm r_1)\cdot(\nhat_{20}+\epsilon_1\Y\times\nhat_{20}+\delta_1\Y)\nl
    +d_{10}\bm B(\bm r_2)\cdot(\nhat_{01}+\epsilon_2\Y\times\nhat_{01}+\delta_2\Y).
\end{align}
Even in the zeroth-order approximation $\bm B(\bm r)\approx\bm B(\bm r_0)$, these terms will not generically cancel, instead leaving
\begin{align}
    \Delta\approx\bm B(\bm r_0)&\cdot\Bigg[\Y\times(\epsilon_0\bm d_{21}+\epsilon_1\bm d_{02}+\epsilon_2\bm d_{10})\nl
    +\left(\delta_0d_{21}+\delta_1d_{20}+\delta_2d_{10}\right)\Y\Bigg],
\end{align}
where $\bm d_{mn}=\bm r_m-\bm r_n$.  In this case, $\Delta=\mathcal O(\epsilon Bd)$ [if $\epsilon_n,\delta_n$ are of the same order].

\bibliographystyle{JHEP}
\bibliography{references.bib}

\end{document}